\documentclass{article}

\usepackage{arxiv}

\usepackage[utf8]{inputenc} 
\usepackage[T1]{fontenc}    
\usepackage{hyperref}       
\usepackage{url}            
\usepackage{booktabs}       
\usepackage{amsfonts}       
\usepackage{nicefrac}       
\usepackage{microtype}      
\usepackage{lipsum}
\usepackage{graphicx}
\usepackage{enumitem}

\title{BEEMA: Braille Adapted Enhanced PIN Entry Mechanism using Arrow keys}

\author{
 Balayogi G \\
  Department of Computer Science\\
  School of Engineering and Technology\\
  Pondicherry University\\
  Puducherry - 605014. \\
  \texttt{balayogistark@gmail.com} \\
   \And
 Kuppusamy K S \\
  Department of Computer Science\\
  School of Engineering and Technology\\
  Pondicherry University\\
  Puducherry - 605014. \\
  \texttt{kskuppu@gmail.com} 
}

\begin{document}
\maketitle
\begin{abstract}
Persons with visual impairments have often been a soft target for cybercriminals, and they are more susceptible to cyber attacks in the digital environment. The attacks, as mentioned above, are because they are visually/aurally exposed to the other sighted users. Visually impaired computer users suffer from secrecy and privacy issues on digital platforms. This paper proposes a mechanism termed BEEMA(Braille adapted Enhanced PIN Entry Mechanism using Arrow keys) to help people with visual impairments. We have studied various security attacks on visually impaired users and proposed a mechanism named BEEMA that provides a rigid braille-adapted text input for people with visual impairments. This mechanism allows users to enter a PIN number on any website that requires a PIN number. The proposed model is implemented as a browser plugin which can be accessed easily. We have conducted sessions with visually impaired users to study the mechanism's performance. The proposed BEEMA model has shown encouraging results in the user study. Resilience of BEEMA against various attacks is also explored in this paper.
\end{abstract}


\section{Introduction} \label{sec1}
The World health organization (WHO) report indicates that 15\% of the world's population is disabled, 285 million are visually impaired, 39 million are blind, and 249 have low vision. Assisting with assistive technology for differently-abled users is essential for them to participate in society.\cite{WHO_2022} In January 2021, the number of active Internet users is about 4.66 billion, 59.5\% of the world population \cite{datareportal_2022}. And the number of users who are accessing the internet through mobile devices and tablet devices is around 4.32 billion, which is about 92.6\% of the active users \cite{statista_2022}. According to the Web Content Access Guidelines (WCAG) guidelines proposed by World Wide Web Consortium (W3C), it is mandatory for all to access the digital ecosystem without any difficulty. \cite{W3C_2022}. Accessible computing is a domain in computer science which tackle and provide solutions, assistive technologies and tools to persons with disabilities. \cite{Deravi2003}\cite{Ismailova2017}\cite{Yakup_2021}\cite{Peker_2022}\cite{Tsekleves2013}\cite{Csontos2021}\cite{Sauer2010}\cite{Balaji_2018}

\subsection{Braille} \label{sec1.1}
Braille is a form of assistive method that used by visually impaired users for learning. There are some braille based keyboards for entering PINs, passwords and more.\cite{Ellis_2020} \cite{Doi_2021} Persons with visual impairments use screen readers softwares such as Non-visual desktop access (NVDA)\footnote{https://www.nvaccess.org/}, Orca, and Job access with speech (JAWS)\footnote{https://www.freedomscientific.com/products/software/jaws/}. In mobile environment, most of the visual impaired users use Talkback and VoiceOver features. The Braille writing system are a group of dots, the presence and absence of the dot represent each letter and digits. The most common form of braille is of 6 dots with 3 rows and 2 columns (2x3). 

Typing PINs covertly is a difficulty faced by persons with visual impairments. Screen reader software will provide audio feedback for every key the user presses on their keyboard. There is a high chance of aural eavesdropping on sensitive information such as passwords and PINs. Without feedback, entering PINs for persons with visual impairments is difficult. Our mechanism focused on presenting an accessible PIN entry system adapted from braille patterns. The proposed mechanism is BEEMA (Braille adapted Enhanced PIN Entry using Arrow keys). 

The major contributions of this article are listed below,
\begin{enumerate}[label=(\alph*)]
    \item To propose a mechanism for accessible and secure PIN Entry for persons with visual impairments titled BEEMA.
    \item To design a browser extension to assist persons with visual impairments in entering PINs.
\end{enumerate}

The remaining of this article is structured as follows: The \textbf{Sec \ref{sec2}} provides the related works. \textbf{Sec \ref{sec3}} discusses about the mechanism and the implementation of this BEEMA mechanism. \textbf{Sec \ref{sec4}} presents the experiments and results. and finally concluded with the future works of this research.

\section{Related works} \label{sec2}
With the increase in the cybercrimes in the world, its important for anyone to be safe in the digital ecosystem. Oftentimes, persons with visual impairments faces many difficulty in having a privacy and secure environment. Though the technology evolves and several assistive technologies have been developed, still the security issues for them are a open ended question \cite{Wang_2020}.

Banerjee et al., \cite{Banerjee_2018} have made some improvement in the PIN and Password entry technique for visual impaired users in mobile platforms and it was designed using grids. The significance of this technique are resistant against shoulder surfing, Man-In-Middle, Data modification and dictionary attacks. And this technique fails when the malicious hacker uses some screen recording attacks on the victim. 

Azad et al., \cite{Azad_2019} have designed a graphical authentication model by combining two popular existing techniques such as: Pass Points and (Press Touch Code) PTC. And this Hybrid PIN entry technique provides better resistant against attacks.

Chitra et al., \cite{Chithra_2021} have designed a mechanism for authentication without the user need to type out the complex combination of letters, digits onto the system for solving CAPTCHA. The end-to-end encryption technique used in this work are resilient against various attacks such as HMM and Fuzzy solver based attacks.

Kamarushi et al., \cite{Kamarushi_2022} have made a huge improvement in the PIN entry mechanism for people with blind and low vision users. This mechanism uses vibration and a timing-based approach to avoid shoulder surfing and aural eavesdropping issues. The significance of this works are vibration-based approach provides better resistant towards shoulder surfing and aural eavesdropping attacks.

\section{BEEMA: Mechanism and Implementation} \label{sec3}
\begin{figure}
    \centering
    \includegraphics[width=0.8\textwidth]{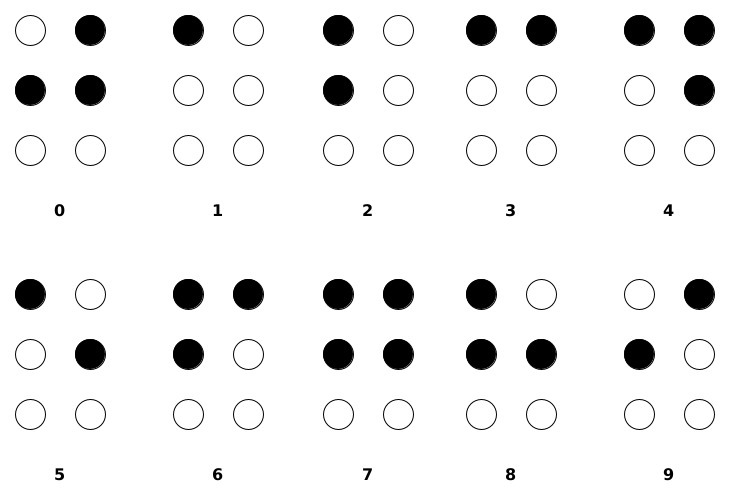}
    \caption{Numerical Digits in Braille form}
    \label{fig:digits}
\end{figure}

This article proposes a mechanism for entering PINs on the web for persons with visual impairments. A browser-based assistive tool that helps the user to enter the PINs onto the web portals or forms with an easy two-key press combination. This technique is named BEEMA, abbreviated as Braille adapted Enhanced PIN Entry Mechanism using Arrow keys. As the name suggests, this mechanism is primarily based on the arrow keys. This proposed mechanism will provide tone-based feedback for every pattern the user press using the arrow keys. The pattern for the arrow keys is designed by adapting Braille, a tactile writing system used by visually impaired people. The numerical digits of the braille are presented in \textbf{Fig \ref{fig:digits}}. 

\subsection{BEEMA Architecture Design} \label{sec3.1}
\begin{figure*}
    \centering
    \includegraphics[width=\columnwidth]{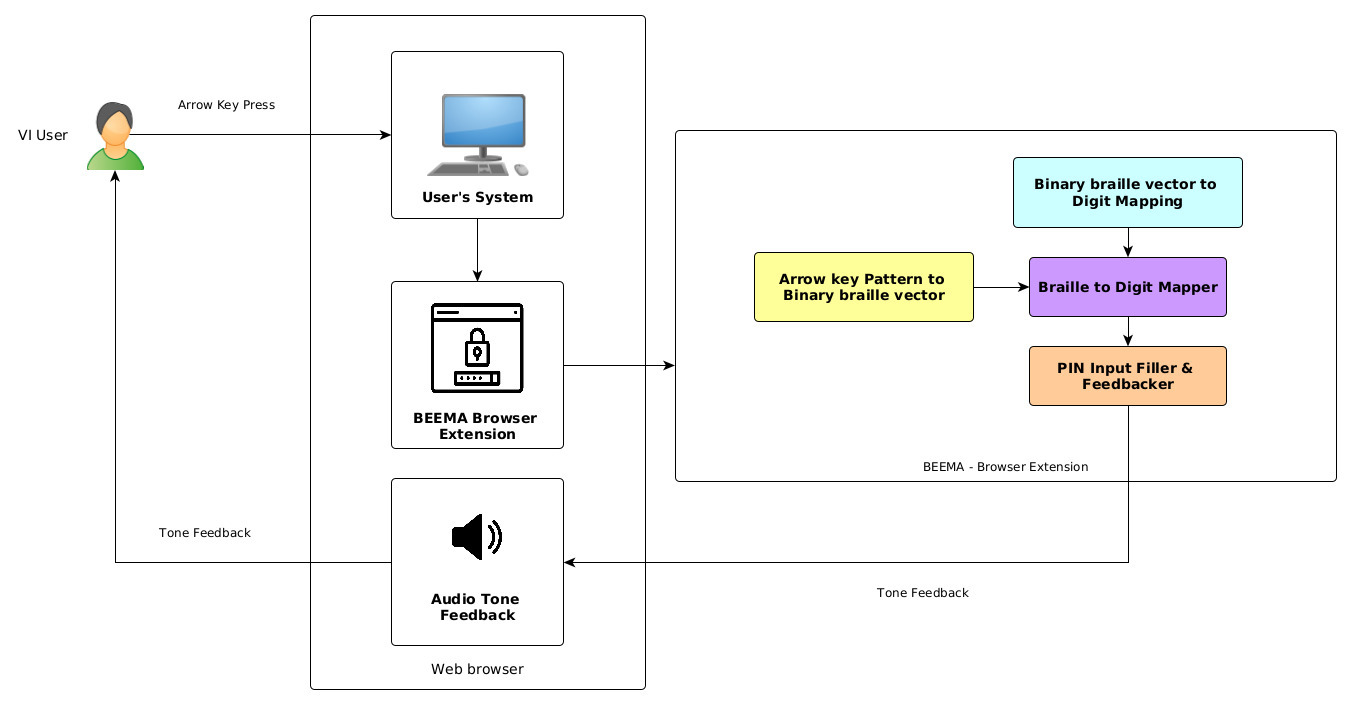}
    \caption{BEEMA Architecture diagram}
    \label{fig:architecture}
\end{figure*}

\begin{figure}
    \centering
    \includegraphics[width=\columnwidth]{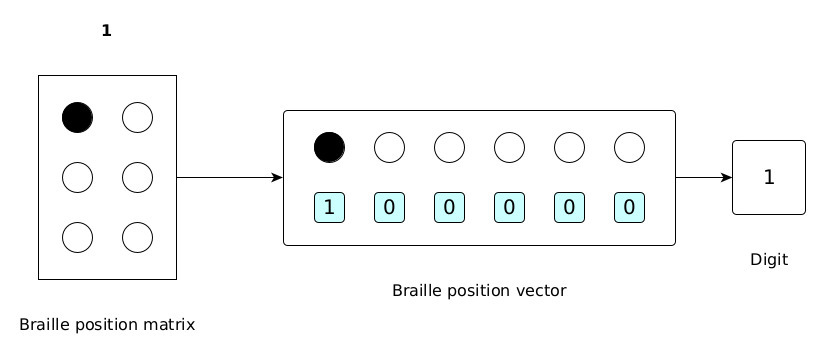}
    \caption{Conversion of Braille letter position matrix to Braille letter position vector and Digit}
    \label{fig:conv}
\end{figure}

As per the architecture of the BEEMA browser extension, it has two core sub-modules, such as Arrow key pattern to Braille vector and Braille vector to Digit mapper. The process of each sub-modules is discussed in this section. In arrow key pattern to binary braille vector generation, pressed keys are captured to construct a binary string of length 6. This vector is constructed with a simple logic of row-wise concatenation as shown in \textbf{Fig \ref{fig:conv}}. Furthermore, in Braille vector to Digit mapper, the corresponding numerical digit of the binary string will be fetched, and the process continues until it reaches the limit of the PIN.

\subsection{Browser extension} \label{sec3.2}
We have proposed and designed a browser plug-in that assists users in entering PINs into web forms and the places where the PIN entry is required. This browser plug-in will assist visually impaired users in entering PINs without explicitly typing out the PINs using a number pad on the keyboard. This BEEMA will read the HTML source code of the user’s web page and find all the $<input>$ tags given with password as their type attribute [type=’password’]. Using the JavaScript code, we have extracted the input tags. We have constructed a mapping between the braille position vector and the numerical digit. Using the “key down” event, we have to identify which key on the keyboard has been pressed. Moreover, the arrow keys are mapped for the generation of string patterns. Where the up arrow key generates 1, the down arrow key generates 0, and the right arrow key will clear the values present in the input boxes. Furthermore, by constructing a string pattern using 1’s and 0’s. Where 1 indicates the presence of the dot and 0 indicates the absence. The process of this BEEMA mechanism is presented in the following steps.

\begin{enumerate}
    \item Initial step of the BEEMA mechanism is to generate a Braille to Numerical Digit mapping as shown in \textbf{Table \ref{tab:mapping}}.
    \item Identify the key press of the arrow keys using the key down event in Java Script.
    \item Using Java Script, Map the up arrow key to generate 1, the down arrow key to generate 0, and the right arrow key to clear the input fields.
    \item Construct a string with a length of n (here, six as we use Braille with six dot representation) using 1's and 0's as per the user's key press.
    \item If the constructed braille vector is present in the Braille to Numerical Digit mapping, then fetch the digit from the mapping and fill it in as the input box value. And provide the user with a tone that denotes each digit. Steps 2 to 5 repeat until it reaches the limit L (here, the limit is a four-digit PIN).
    \item If the constructed braille vector is not presented in the Braille to Numerical Digit mapping, they give the user a different tone.
\end{enumerate}

\subsection{Implementation} \label{sec3.3}
\textbf{Fig \ref{fig:architecture}} shows the complete architecture diagram of the BEEMA mechanism. This BEEMA mechanism uses arrow key combinations to enter various patterns of binary string that denote braille digits. Whenever the user makes the binary pattern in the braille to numerical digit mapping, the extension will give tone-based feedback so that the user can understand that they have entered a PIN. The tone will differ for each digit, so the user can easily distinguish between the tones. The process of the BEEMA browser extension is discussed in \textbf{Sec \ref{sec3.2}}. The functionality was implemented using JavaScript, making the PIN entry faster on the client side.

\begin{table}
    \centering
    \begin{tabular}{c c}
    \hline \\
        \textbf{Braille position vector} & \textbf{Numerical digit} \\ \\
    \hline \\
        011100 & 0 \\
        100000 & 1 \\
        101000 & 2 \\
        110000 & 3 \\
        110100 & 4 \\
        100100 & 5 \\
        111000 & 6 \\
        111100 & 7 \\
        101100 & 8 \\
        011100 & 9 \\ \\
        \hline \\
    \end{tabular}
    \caption{Braille String Pattern to Numerical digit mapping}
    \label{tab:mapping}
\end{table}

\section{Experiments and Results} \label{sec4}
\begin{figure}
    \centering
    \includegraphics[width=\columnwidth]{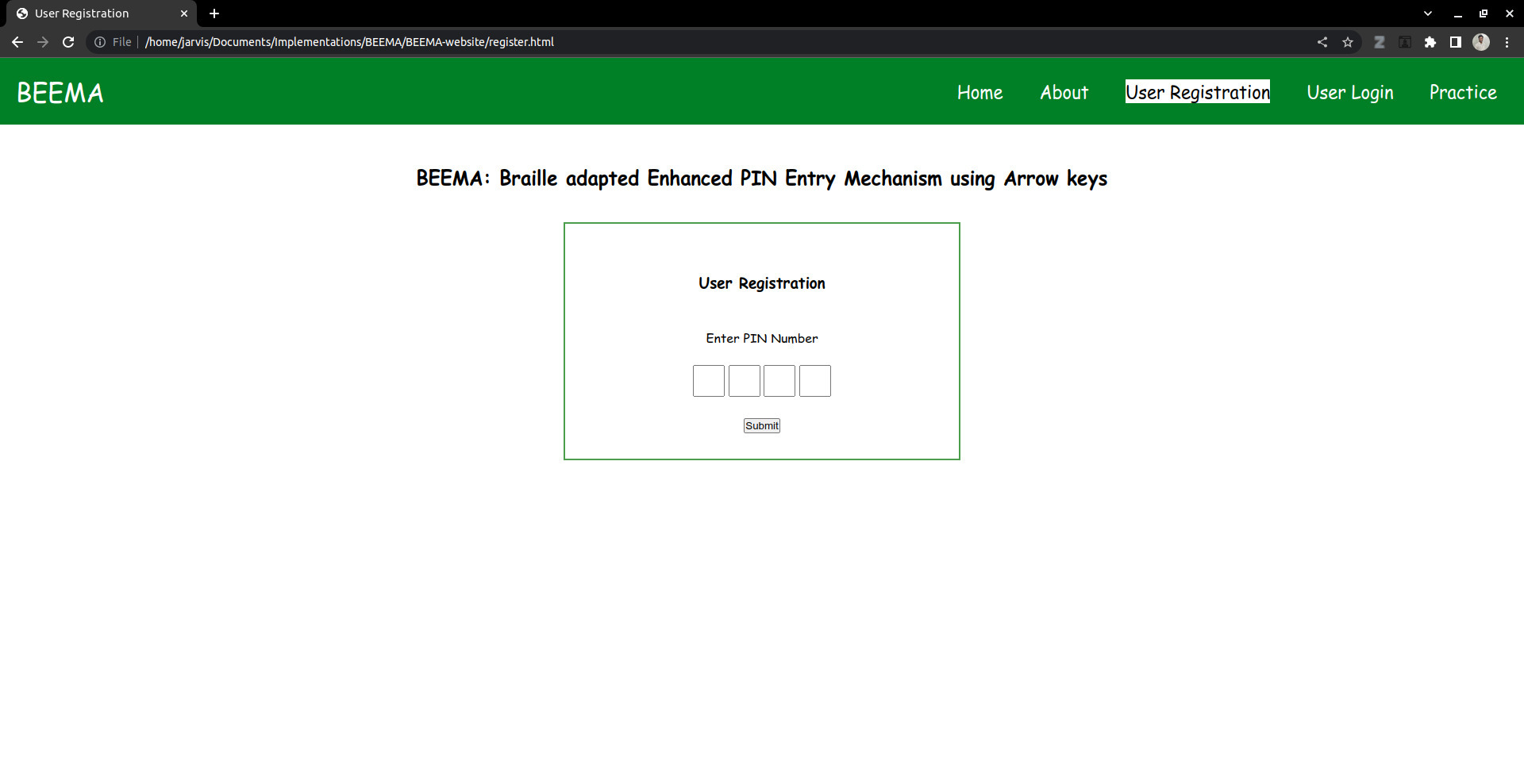}
    \caption{Four digit PIN Registration page}
    \label{fig:register}
\end{figure}

\begin{figure}
    \centering
    \includegraphics[width=\columnwidth]{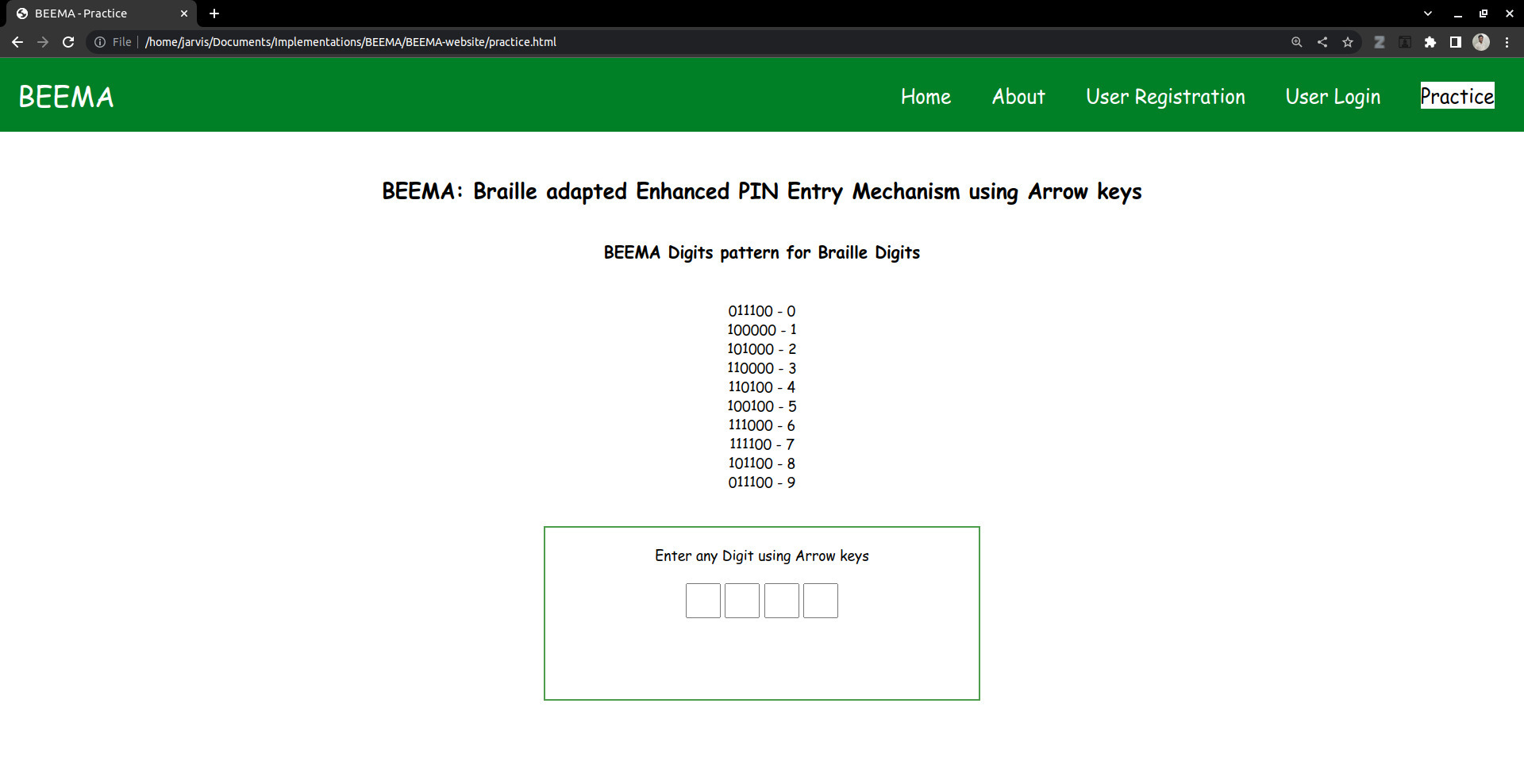}
    \caption{Practice page}
    \label{fig:practice}
\end{figure}

To evaluate the performance of this BEEMA browser extension, we have designed a demo website with a PIN entry interface using HTML and CSS, as shown in \textbf{Fig \ref{fig:register}}. As this mechanism requires remembering binary patterns, we have designed a prototype web page with a PIN interface and told them to use it. With the help of Java Script, we included a script that will provide audio feedback to the user (only for practice purposes) as shown in \textbf{Fig \ref{fig:practice}}.

\subsection{Usability analysis} \label{sec4.1}
\begin{figure}
    \centering
    \includegraphics[width=\columnwidth]{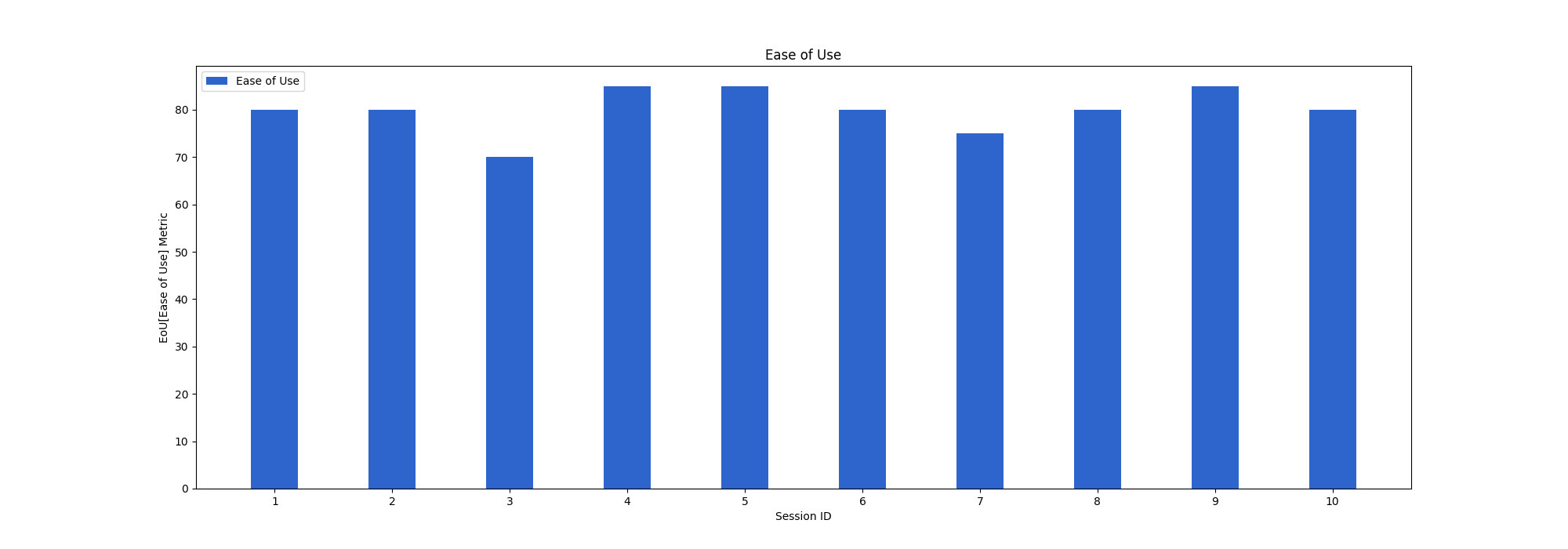}
    \caption{Ease of Use of BEEMA mechanism}
    \label{fig:eou}
\end{figure}

We conducted several sessions with visually impaired users and asked them to use the BEEMA browser extension and provide feedback.\footnote{https://github.com/BalayogiG/BEEMA.git} The usability analysis is carried out in two components:  \\
(i) Ease-of-Use (EoU), which is used to quantify on a Likert scale\cite{Jamieson_2004} of 1 to 5, where 1 represents the least satisfaction, and 5 represents maximum satisfaction in our BEEMA mechanism. The feedback collected from the visual impaired users and it shows that the 80\% of the users are certain that the BEEMA mechanism is easy to use as shown in \textbf{Fig \ref{fig:eou}}. 

\noindent (ii) Time is taken to enter a PIN and the average time taken to enter four digit PIN is approximately 0.22 seconds. 

\subsection{Security analysis} \label{sec4.2}

\subsubsection{Shoulder surfing attack} \label{sec4.2.1}
Shoulder surfing depicts a circumstance in which an attacker may physically observe the device's display and keyboard to get sensitive data. It is an effective approach for gathering information in active regions as it is quite simple to stand next to someone and observe them while they fill out a form, enter their PIN in an ATM, or pay with a credit card. The attacker must be close to the victim, making this a rather uncommon kind of assault. These types of attacks can also be performed from a great distance utilizing binoculars or other vision-improving gear. Technically, shoulder surfing is a sort of social engineering. It essentially indicates that an unauthorized third party may observe a monitor and/or any private information shown on an electronic device. It's very uncommon for people to be concerned about their privacy while working in public places like cafes or open offices, where customers, co-workers, and passers-by may be able to spy on their screens as they pass by \cite{BOSNJAK_2019}.

The danger of shoulder surfing may be significantly avoided using simple, cost-effective measures. A simple approach to prevent shoulder surfing would be to sit with your back against the wall. In this manner, you restrict the capacity of others to glimpse your screen or view your data. Using a privacy screen, you may also safeguard your computer from shoulder surfing. Unfortunately, these cheap displays may hinder your everyday use. Additionally, shoulder surfing danger is not exclusive to public spaces. Frequently, attackers want to acquire visual access to an individual's computer display while the individual is at their typical workplace and unaware. For instance, while guests explore the office, they may quickly peek at an individual's displays as they go from one area to the next. 

Many individuals mistakenly feel they are protected from malevolent intentions at the workplace, but suppliers, onsite clients, customers, and even colleagues should be seen as potential privacy hazards. Safeguarding yourself against intruders by avoiding shoulder surfing is something you should always do, whether at home, at the office, or in a public place. As this BEEMA browser extension uses only two keys to generate every digit, it is difficult to see or understand what they are typing on the screen. The absence of hand movement provides an additional advantage in resisting the shoulder surfing attack.

\subsubsection{Screen shot attack} \label{sec4.2.2}
Screen shot attack captures a snapshot of the hacked system's screen to obtain data about the computer device. For intruders, the exposure of any confidential material is a prime target. This includes financial data, login details retained insecurely (on a notepad, Word document, etc.). A screenshot may be taken once using screen capture, or it can be configured to take snapshots at predetermined intervals. The hackers are aware that the screen shot attack approach may provide critical information in an attack campaign, regardless of how it is performed. In a screenshot attack, the cybercriminal intends to capture the display of the PINs to gain access to the account. Many websites use the password masking feature to hide the passwords from view, and This attack has been reduced. Even if they capture the screen, it does not reveal the PIN to the attackers.

\subsubsection{Key logger attack} \label{sec4.2.3}
Keylogger attacks are the first cyber hazards. It scans and records keystrokes and can identify trends to make it simpler to uncover passwords. In other words, a keylogger is software that may capture and notify a user's behavior while they communicate with the system. Keylogger is an abbreviation for "keystroke logger," referring to the fact that these programs monitor your activity by capturing your keystrokes as you write.

Malevolent hackers often deploy keyloggers for evil objectives. The idea of someone following your every move may be unsettling. However, there are valid or at least legal applications for keyloggers, including parental monitoring of children's internet activity and employer monitoring of employees. They are often disseminated by viruses, USB drives, software, and hardware flaws. Before opening and executing a file, verifying that the activity will not propagate a keylogger is essential.
    
\begin{figure}
    \centering
    \includegraphics[width=\columnwidth]{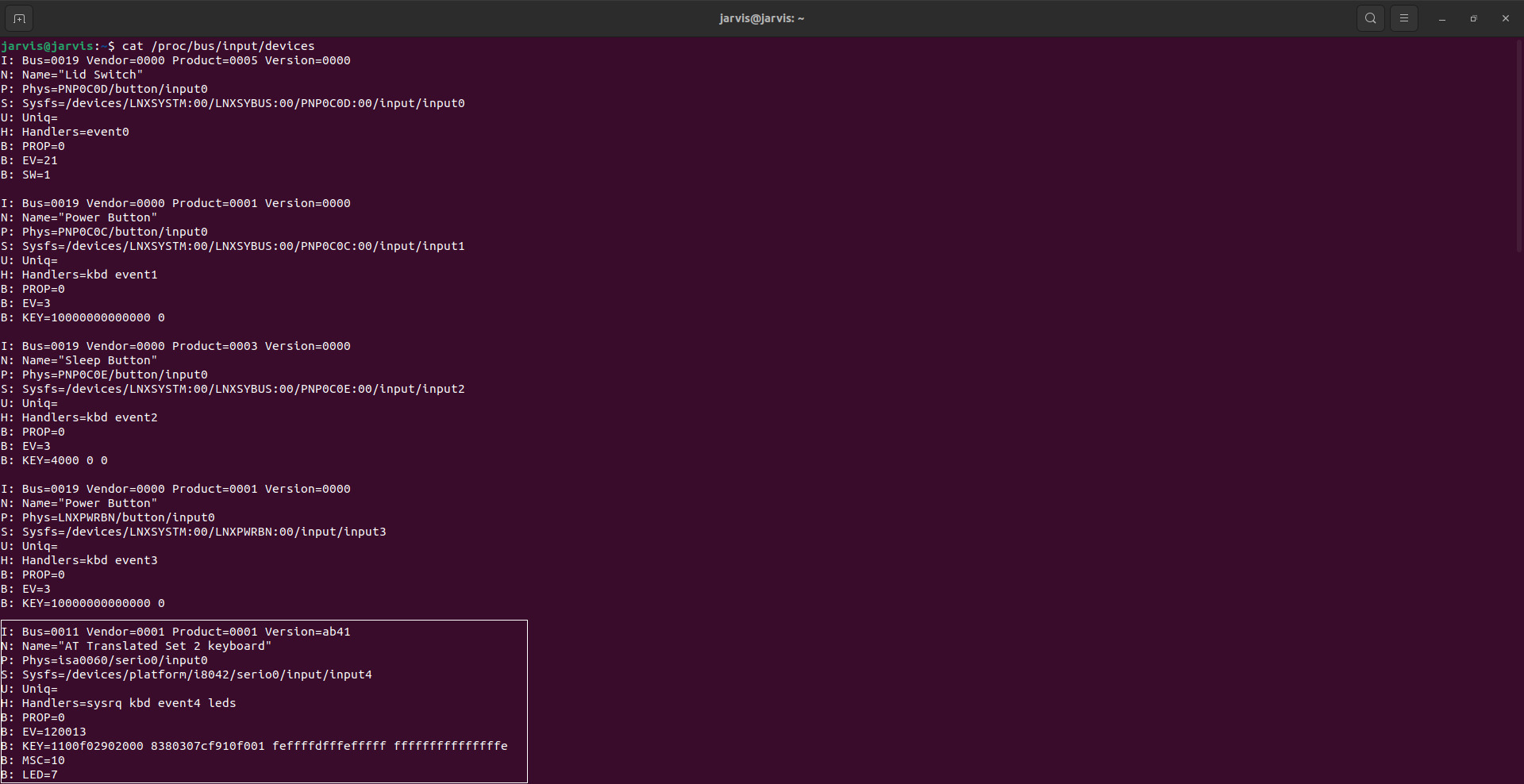}
    \caption{Input device identification for identifying keyboard}
    \label{fig:input}
\end{figure}

\begin{figure}
    \centering
    \includegraphics[width=\columnwidth]{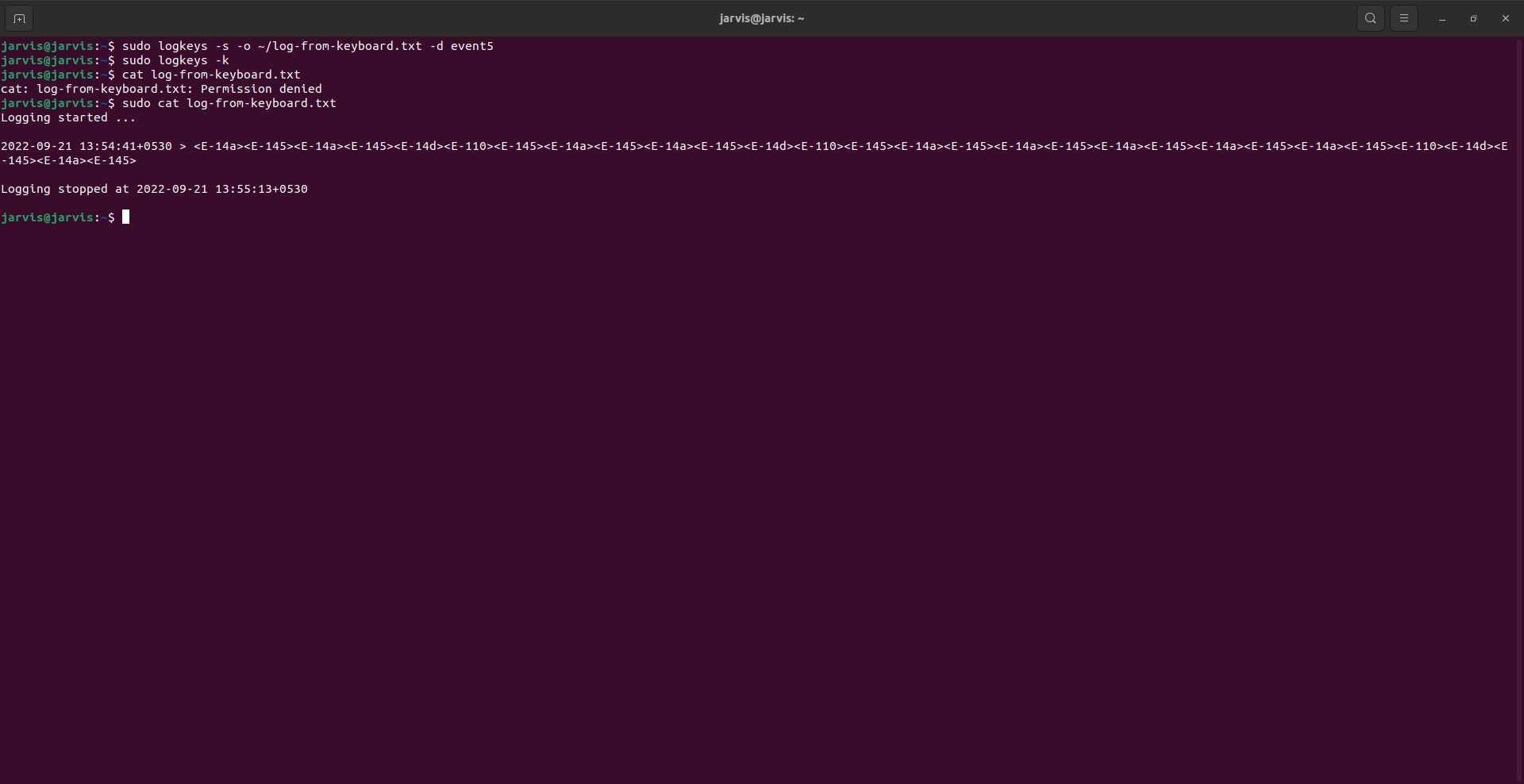}
    \caption{Key logs captured using logkeys}
    \label{fig:keylogs}
\end{figure}

As this BEEMA browser extension is working based on the keyboard keys. There are more chance of being attacked with key loggers. So we used Keylogs\footnote{https://github.com/kernc/logkeys.git} key logger to capture and see the key logs. To capture the key logs we have identify the input device that is keyboard. In order to capture the keyboard key presses, we have to identify the input device ID. So we use the Linux command to list input devices as shown in \textbf{Fig \ref{fig:input}}. And the captures keys are displayed using $cat$ command as shown in \textbf{Fig \ref{fig:keylogs}}. From the analysis, the hacker can only know what key is pressed, and they can not be able to conclude which PIN the user is using.

\section{Conclusion} \label{sec5}
In this article, we have proposed a mechanism for entering a PIN into the web forms for persons with visual impairment. Often, banking websites are provided with the PIN-based entry for OTP pins for transactions. We have designed a browser extension to assist visually impaired users in entering a PIN on the web. This browser extension utilizes the keyboard and is resilient to key logger attacks. We have conducted various sessions with visually impaired users and asked them to use and provide feedback in terms of usability. We have received positive feedback from visually impaired users. From the feedback, 80\% of users expressed that it is an easy-to-use (EoU) metric, and the time taken to enter the PIN is also improved. Furthermore, this mechanism is resilient against shoulder surfing attacks. The future research direction for this research will be an expansion to expand the password entry by incorporating letters, digits, and symbols. So visually impaired users can use an easy way to enter the password. 

\bibliographystyle{unsrt}  


\bibliography{references}

\section*{Recognition} 
This paper was presented at the International Conference on High Performance and Intelligent Computing (ICHPIC) on 8th and 9th December 2022. only abstract is printed in the conference proceedings.

\end{document}